# Shannon Revisited

*Considering a More Tractable Expression to Measure and Manage Intractability, Uncertainty, Risk, Ignorance, and Entropy*

Gideon Samid
Electrical Engineering and Computer Science Department
Case Western Reserve University
Gideon.Samid@case.edu
13 Tapiola Ct., Rockviile, MD 20850, USA
Ph: 1-301 424 7990  FAX: 1 301 424 4147
Professor (Adjunct), University of Maryland gsamid@faculty.umuc.edu

*Abstract: Building on Shannon's lead, let's consider a more malleable expression for tracking uncertainty, and states of "knowledge available" vs. "knowledge missing," to better practice innovation, improve risk management, and successfully measure progress of intractable undertakings. Shannon's formula, and its common replacements (Renyi, Tsallis) compute to increased knowledge whenever two competing choices, however marginal, exchange probability measures. Such and other distortions are corrected by anchoring knowledge to a reference challenge. Entropy then expresses progress towards meeting that challenge. We introduce an 'interval of interest' outside which all probability changes should be ignored. The resultant formula for Missing Acquirable Relevant Knowledge (MARK) serves as a means to optimize intractable activities involving knowledge acquisition, such as research, development, risk management, and opportunity exploitation.*

# 1.0 Introduction

In 1948 Claude Shannon proposed his information content (or entropy) formula to characterize the level of knowledge, or alternatively, level of ignorance associated with a situation where several 'states' or 'outcomes' are possible, and their probabilities are duly assigned [Shannon-48, 49, Brilloun-62]. For *n* states with probabilities: $p_1, p_2, \ldots, p_n$, Shannon entropy, *H*, is given as:

$$H = -\sum_{i=1}^{n} p_i \log(p_i)$$

Shannon has further shown [ibid] that his formula is the only analytic expression that would comply with what he considered as logical requirements: (i) continuity with the probability ratings, (ii) monotonicity with n, and (iii) the condition:

$$H (1,2,...n) = = -\sum_{i=1}^{n} p_i log(p_i) + \Sigma\, p_i H(i)$$

for a case where *n* candidates compete to be '*chosen*' and each candidate can be further divided into sub-elements, and so on, iteratively. The original use of the formula was in communication, coding, and cryptography, but it has since spread to a variety of applications where uncertainty is expressed through rated probabilities of competing outcomes. The range of applications can be categorized as (i) action oriented, and (ii) analysis. In the latter one usually applies the *maximum entropy principle* to describe uncertain situations, and track natural behavior, be it environmental parameters, evolutionary pathways, DNA dynamics, etc. [Mazaheri-10, Telesca-08, Liu-09, Okada-08, Zhong-09, Shahkooh-09]. In the former one makes use of the measures of entropy to optimize a target activity.[ Balta-08, Gang-09, Golic-08, Jablonowski-03,95, Li-97, Quero-10,Shakooh-08]. In both categories Shannon's entropy achieved a status of prominence that tends to overshadow its inherent arbitrariness. Shannon's entropy satisfies its self imposed conditions (see above), and is remarkably similar to the thermodynamic entropy formula deduced from statistical mechanics. It is also the only analytic expression for its purpose. So being, it lends itself to elegant mathematical analysis of maximum entropy subject to a set of conditions and constraints [Chen-08]. Shannon's entropy also is very sharp at its boundary conditions. It reflects convincingly the state of total ignorance and the state of total knowledge. Albeit, the passage from the former state to the latter is 'generic' in as much as it does not relate to the particular aim of the action that leads to knowledge acquisition and to reduced entropy. The probability distribution at any interim state may be more beneficial to one action than to another. Shannon though, does not allow for such distinction. In revisiting Shannon one would explore an alternative expression for entropy such that only changes in probabilities that impact the target action will be accounted for.

Shannon entropy seems a perfect instrument to measure the information content difference between, say, a proper English text, and an equal size string of random characters. In part this is because the competing options (choice of letters) is unambiguous. When Shannon's formula is applied to situations where the options themselves are arbitrary, then it leads to some strange results.

Let's consider a search for a missing person where our level of ignorance is maximized -- we have absolutely no knowledge where this person is – except that he is somewhere on the planet. This case can be formulated through a binary distinction: is that person in America, or elsewhere? According to that definition the entropy of the situation is H=1, because our complete ignorance dictates that the chance for that person to be in the US vs. his chance to be elsewhere is: $p_{America}= p_{elsewere}= 0.50$. Albeit, we can rephrase the case by asking: is the missing person in North America, in South America, in Euro-Asia, in Africa, or in Antarctica? Our level of ignorance or knowledge about the case has not changed, however, Shannon's entropy will now indicate H=2.32 because our total ignorance will be expressed as:

$p_{North\ America}= p_{South\ America}= p_{Euro-Asia}= p_{Africa}= p_{Antartica}= 0.20$   This is an example where the definition of states of outcome is arbitrary, and this arbitrariness spills over to the measure of the entropy of the situation, discrediting the applicability of the result.

Let's explore a different case : Two investments X and Y are under consideration. Each investment is associated with two scenarios: $X_1$, $X_2$ and $Y_1$ and $Y_2$. In case (a) the calculated chance for the four scenarios is: $X_1$=10%, $X_2$=40%, $Y_1$=10%, $Y_2$=40%. In case (b) the probabilities are: $X_1$=40%, $X_2$=40%, $Y_1$=10%, $Y_2$=10%. Shannon's entropy in both cases is 1.72 Yet, investment X is much more attractive in the (b) configuration.

For a third example consider the generic question of value (scalar) estimate. Our objective is to find the true or right value of parameter X, to be denoted by $x_{true}$. In a typical estimating situation one develops a probability curve *p(x)* such that the chance for $x_{true}$ to be within the interval x=a to x=b is given by:

$$P(a,b) = \int_a^b p(x)dx$$

One could compute Shannon's entropy associated with a given *p(x)* by converting the curve into a histogram. One then faces the question of the size of the intervals of the chosen histogram. The smaller the interval, the more distinct options are defined, and the higher the entropy value associated with the same situation of ignorance or knowledge. When Max Planck faced a formally similar dilemma in analyzing black body radiation, he solved it by looking for a natural size interval -- and thereby jump started the new physics of quantum mechanics. Such "natural interval" is elusive for our case, but without it one faces the Planck dilemma, namely:

$$\lim_{I \to 0} H(estimate) = \infty$$

Where *I* is the interval size the columns of the histogram. It is noteworthy that this inconvenience of 'running to infinity' was formally resolved by conveniently adopting a Shannon variety formula:

$$H = \int_{-\infty}^{+\infty} p(x) \log(p(x))\, dx$$

Only that this creates a singularity point with the discrete case. Let p(x) be a uniform flat line stretching between two arbitrary points x=a to x=b. If the case is converted to a histogram with column width (interval) I, then the discrete entropy will compute to $H = \log \frac{b-a}{I}$. While the continuous entropy will evaluate to $H = \log(b - a)$ Which is the "true" entropy?

Claude Shannon himself appeared to have been taken aback by the plethora of applications that his formula was used for, saying: "*workers in other fields should realize that the basic results of the subject are aimed in a very specific direction*",[Shannon-56]. Recent publications have also called for re-examination of the old formula [Wang-09, Benjun-09, Wang-09b, Ding-07, Titchener-00]. As early as 1961 A. Ré´nyi (Renyi-61, Brechner-07) offered more "playroom" by introducing an α-parameterized expression:

$H_\alpha(x) = \frac{1}{1-\alpha} \log \int_{Range} p^\alpha(x)dx$  or for the discrete form:  $H_\alpha(x) = \frac{1}{1-\alpha} \log(\sum_{i=1}^{i=n} p(x_i)^\alpha)$

For α >0, and α≠1, which collapses to Shannon's formula for α=1. Tsallis [Tsallis-01] offers his variety:

$$H_\alpha(x) = \frac{1}{\alpha - 1}(1 - \sum_{i=1}^{i=n} p(x_i)^\alpha)$$

Golshani offers a detailed review of the various Shannon-replacement propositions and their common use in computing 'maximum entropy'[Golshani-10]. Shannon entropy underwent several more reinterpretations: the probability ratings were replaced by fuzzy memberships in various classes, (Zheng-08, Ding-07); "Discrete entropy" is discussed by Amigo [Amigo-07], and "permutation entropy" is proposed by Bandt [Bandt-02] . Occasionally Shannon entropy is used to measure dispersion in lieu of standard deviation [Chen-09]. These propositions do not focus on replacing Shannon as a means to appraise and quantify the knowledge gap between a state of uncertainty and its corresponding state of certainty, and they do not resolve the issues raised in the examples above. The discussion below, is addressing this unattended aspect of the ingenious Shannon idea. It focuses on the difficulties exemplified by the previous illustrations, and on the need to quantify the knowledge gap between two progressive states of an uncertain situation.

## 2.0 "Knowledge" Revisited

The concept of *knowledge* has been mentioned, used and discussed in circles of philosophy and science for as long as these disciplines existed. Yet, a quantitative measure thereto is still elusive.

One fundamental reason for our inability to quantify knowledge is that the situation appears *'bottom less'*: namely, that we are hard pressed to define a state of zero knowledge. In fact, it appears that our state of knowledge at any given time is an accumulation of wisdom, lessons, and impressions that have been preserved from our single-cell evolutionary ancestors by the magnificent process discovered by Charles Darwin. We think, know, are cognizant and aware, all with our present day brain that has been painstakingly constructed over millions of years, gradually endowing us with the ability to recognize patterns, to note distinctions, spot similarities, exercise logic, and practice reasoning. Being "*bottom less*" means that we face a daunting challenge when we endeavor to quantify the sum total of knowledge that we possess at any given state. Alas, much as we could measure temperature gaps before we were able to spot the point of zero temperature, so we might opt to measure knowledge while still suffering from the bottom-less state.

We may then endeavor to quantify the "knowledge gap" between two well defined states. And in order to succeed, these two states will have to be very well defined indeed. To achieve the necessary clarity we shall build a limiting model to work with. The model will feature two elements:

- ✓ a knowledge seeker (Kseeker, KS)
- ✓ a reference challenge (RC)

Every piece of data or information may be regarded as knowledge, and the quantity of that knowledge depends directly on the purpose for which that data serves. If I wish to find out the age of the universe then the list of items in my car glove-box constitute zero knowledge, but if my purpose is to find the car's registration then the same information reflects a great deal of relevant knowledge. Say then that in order to map information to knowledge it is necessary to define a purpose. And that is where the reference challenge comes into play. We shall climb down from the lofty ambition to measure knowledge per se, and endeavor to measure it in reference to satisfying a reference challenge, or say in reference to solving a given problem. Additional body of research related to the work reported here is suggesting that knowledge is not necessarily an objective quantity that exists a-priori in the universe, exposed to some, obscure to others [Samid-07]. It might well be that *knowledge is generated in response to seeking it*. But at any rate matters have more clarity if we refer to a particular knowledge seeker, and ask ourselves how much knowledge the seeker possesses. And since, as we argued, it is at present impossible to define the state of zero knowledge, then we must limit ourselves to finding relative knowledge, or better, define *'missing knowledge'*. If the knowledge seeker at state A cannot solve the reference challenge and at state B he does resolve it, then we can attempt to measure the knowledge gap between these two states. That gap is either the knowledge missing at state A, or the knowledge acquired in moving from state A to state B. Where *missing* and *acquired* refers to the particular knowledge seeker and the particular reference challenge.

In order to express knowledge states for an unsolved challenge we resort to a generic problem solving model whereby a problem is associated with N fully specified solution options, and one of these options represents the solution to the challenge. "Fully Specified" solution options are options that can be exercised using the knowledge at hand, with no need to acquire more knowledge. The N option may be grouped into sets, which may be grouped again and again, defining a tree structure of options. 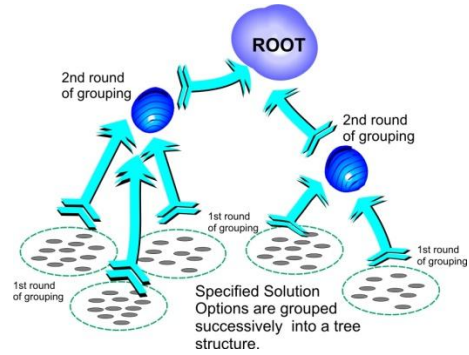 The process of resolving a challenge involves first a high level identification of the right tree branch, then a lower level identification of the right branch, on and on until the fully specified working solution is identified. At any given state in the solution seeking process the solution seeker is aware of n ≤ N solution options at various degrees of specificity. We can now define a "stage solution" for the reference challenge as identifying which of the n identified solution options is the right one (contains the fully specified solution pathway). We shall now focus on the state of knowledge with respect to specifying the stage solution, to be defined as the stage challenge. The n identified solution options can be rendered complete and comprehensive, by insuring that one of them is defined as "another" or "else" with reference to all the other identified explicit solution options. With reference to the stage reference challenge, the solution seeker can identify his, her or its knowledge state by the set of n probabilities rating for the n competing solution options. The probability of each solution option will represent its probability to be the right one.

For any stage solution reference, the state of zero knowledge will be expressed as an even uniform probability distribution, namely $p_i=1/n$ for $i=1,2,...n$. And the state of full knowledge will be expressed as: $p_i(i \neq j)=0;\ p_j=1$. The in-between states will be represented by a probability distribution morphing from the state of zero knowledge to the state of full knowledge. 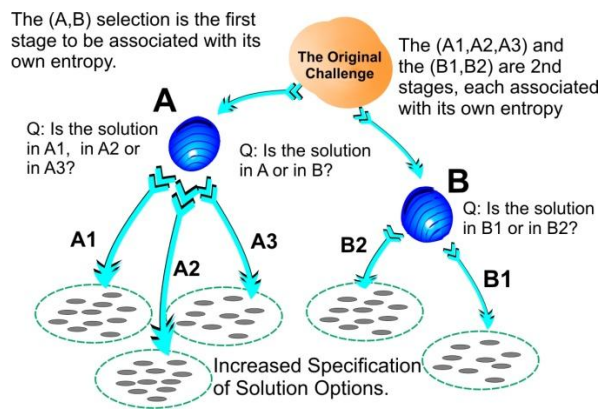

We now turn to define the state of "problem solved". Certain problems have a binary solution status: they are either solved or not solved. Such are "lock and key" problems, like in cryptography. The key either works, or it doesn't. Such are search problems where the object of search is well defined and is either found or not found. Alas, the majority of problems are associated with a range of solution states. If we wish to estimate the cost of something, then specifying a narrow enough range (as opposed to a single accurate value) amounts to a satisfactory solution, and if that range or interval is a bit larger, it would amount to a partial solution. It is similar for construction challenges: while a certain construction will constitute a perfect solution to a given problem, a different construction will serve as an 'almost perfect' or 'good enough' solution. This reality suggests that knowledge quantification faces not only a bottom-less challenge but also a "top-vague" challenge. Only that the latter is much easier to address. We do so through the concept of interval or domain of interest. We define this domain by introducing two concepts: (i) the interval of indifference (IOI), and (ii) the interval of futility (IOF).

A problem may be solved to such an advanced degree that any further refinement will not be of any help, or interest. If we wish to find the weight of an object, then we can always refer to some $\pm\Delta$ weight that is so narrow that we really don't care where in the $2\Delta$ range the "*true*" weight lies. This $2\Delta$ range is the interval of indifference, or IOI.

On the opposite end a problem might be associated with some level of knowledge, alas, of such low measure that all that knowledge is essentially useless, and it would provide no added value by possessing it. If we wish to estimate the cost of an object and we managed to ascertain that the cost is between, say $10 and a $1,000,000 then, we are no better off in comparison to an interval specified between $1 dollar and $10 million. Every interval larger than the *Interval of Futility,* IOF, is as worthless as the IOF, and that is its definition.

In the base (binary) case we will have IOI=IOF. Namely, there is only one interval of interest - Any smaller interval, or a larger interval, is of no interest. That is the case of finding or not finding an object of search. That is also the case in searching for a proper cryptographic key. All non working keys are equally useless. In the general case, where IOF > IOI we have a range of interest between IOI and IOF (e.g. estimating a scalar value). When such a range exists the question of solving a problem is getting a bit more complicated. Solving a problem with respect to any interval of interest I (where IOI ≤ I ≤ IOF), is a partial solution which may qualify as a terminal state. In other words, a problem may be solved with the assumption that a given I value is indeed both the IOI and the IOF (I=IOI=IOF), and one should measure the knowledge gap between a reference state J and I. Let us designate the in between intervals as $I_2$, $I_3$,..:

**IOI, $I_2$, $I_3$, ....... $I_{s-1}$, IOF**

We should consider a situation in which the knowledge-seeker has acquired new information or knowledge that would make it possible to solve the challenge on the basis of some $I_i$ without affecting the solution ability versus any other interval of interest. Because such a situation is possible, it is necessary for us to mind the knowledge gap between any reference state J on one hand, and any of the s intervals of interest, on the other hand. In other words, it would not be sufficient to mind the gap between a

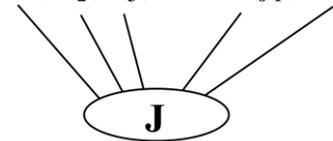

Reference case J has a different knowledge gap vs. different intervals of interest.

reference state J and the IOI because it would miss knowledge that would make it easier to solve the problem with respect to an interval I >IOI.

The above point may be viewed in terms of validity vs. utility. The validity of a statement reflects its likelihood to be true and correct. The utility of a statement reflects its usefulness for the purpose at hand – *if true*. A statement of knowledge with respect to the IOI is very useful, if true. A similar statement with respect to the IOF is quite useless. In general there exists a tradeoff between validity and utility. Validity may be increased for the price of reduced utility. E.g. the statement that the cost of an object is between 1$ to 10$, has more validity than the statement that its cost is between $4 and 5$. But the larger interval may be less useful. The range between IOI and IOF is the range of utility. This view of knowledge does not imply that any change in the probability function corresponds to acquiring more relevant knowledge (in contrast to Shannon's formula). If the probability function changes in such a way that it does not become easier or more likely to solve the reference challenge, then such changes are knowledge-neutral with respect to the reference challenge. Such changes could have been useful, if the knowledge seeker was focused on another challenge perhaps. In fact, we define relevant knowledge as knowledge that reduces the knowledge gap between a current state and at least one of the intervals of interest in this case.

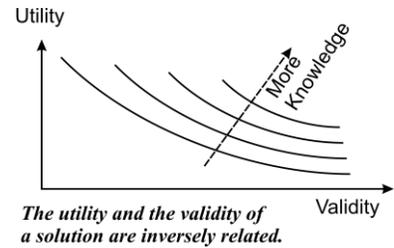

*The utility and the validity of a solution are inversely related.*

If we designate as $\Delta k_{J-I_i}$, the knowledge gap between state J and state $I_i$, and let $K_J$ represent the knowledge state associated with state J, then per our discussion above:

$$K_J = f(\Delta K_{J-IOI}, \Delta K_{J-I2}, \ldots \Delta K_{J-IOF})$$

where the function $f$ remains to be determined. This equation expresses the notion that any change in the knowledge gap between the present state J and any state of interest I (between IOI and IOF) is a change in the total missing solution knowledge for the problem or challenge at hand, and as such it is also a change in the uncertainty, the risk, the opportunity, etc. of the case in point. That is the measure we aim to quantify.

# 3.0 Knowledge Quantified

Knowledge issues may be classified as: (i) discovery, and (ii) construction. The former refers to finding some missing information, and the latter refers to action, and steps to be taken to build, construct an object for a given purpose. These two categories resist a clear distinction because one could think of building a device with which to discover a desired piece of information, and alternatively one could think of discovering the information needed to construct the needed contraption. So we could replace the above distinction with one that is a bit less overlapping: *discrete knowledge versus knowledge of a continuum*. In the first case, the simplest model will be a choice of one among a given finite number of choices, and in the latter the quest will be to point to a sufficiently small zone in a given continuum so that the pointed-to zone adequately represents the required knowledge. In the simplest case we will consider a one dimensional scalar mapped on a continuous line, and the target zone will be an interval on that line. We could later augment this model to an *n*-dimensional zone (metric spaces) [Samid-09]. As to knowledge of which selection to make among discrete choices, we could further augment the choices into a tree structure, and address the nodes of the tree.

### 3.1 Estimating a Continuous Variable
Let a continuous variable *x* be a property of a given object, where it assumes the values $x=x_0$. We shall consider the existence of a *knowledge-seeker* who is trying to find out the value of $x=x_0$, and we shall ask ourselves what is the *amount of knowledge* needed to ascertain the value of $x_0$, and of which the knowledge-seeker is still ignorant (at any given state before actually solving the problem). The imagined knowledge seeker, at any given state of knowledge, may be associated with a probability function *p(x)* that is used to express the probability of $x_o$ residing in a given interval on the continuous line. We can readily define the boundaries of knowledge: total ignorance and total knowledge. The first case will be expressed via a horizontal line parallel, and almost adjacent to the x-axis line, while the latter will be written via a delta function centered around $x_0$.

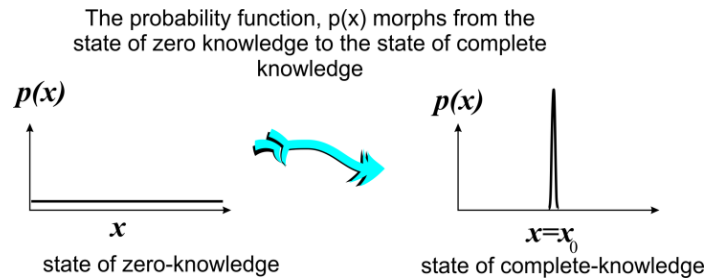

As the knowledge seeker becomes more knowledgeable his probability function *p(x)* morphs from the total ignorance state to the total or complete knowledge state.

In this setting the knowledge seeker may introduce the IOI and the IOF as defined above, with an added specificity that these intervals are expressed as a section on the continuous *x* line. Since the IOI can be set to a small-interval as desired, and the IOF to a large-interval as desired, there is no loss in generality through the use of these two limits.

For any given point on the *x*-axis, $x=x'$, and for any state of the probability curve *p(x)*, and with reference to any given interval of interest I we may compute the probability for $x=x_0$ to be between *x'-0.5I* and *x'+0.5I* as follows:

$$P(x', I, p(x)) = \int_{x'-0.5I}^{x'+0.5I} p(x)dx$$

For some value of $x'=x^*$, the value of $P(x^*, I, p(x))$ will be maximum. We designate that value as $\pi(I, p(x))$[1]. If it so happens that the IOI to IOF interval collapses into I then $\pi(I, p(x))$ will represent the chance for the interval $(x^*-0.5I)$ to $(x^*+0.5I)$ to include $x_0$. And since that chance is the highest for $x=x^*$, then $x^*$ will be the best guess for $x_0$.

For any I' > I, we can write: $1 \geq \pi(I', p(x)) \geq \pi(I, p(x))$ for a given $p(x)$. In terms of $\pi$ the two boundary cases appear as below:

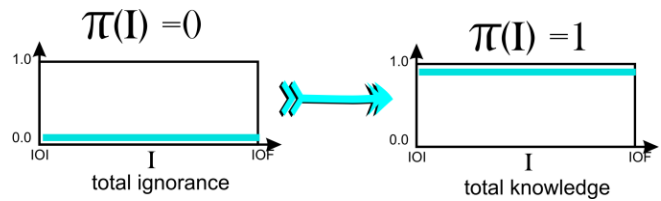

And the $\pi$ curve for progressive knowledge states look like:

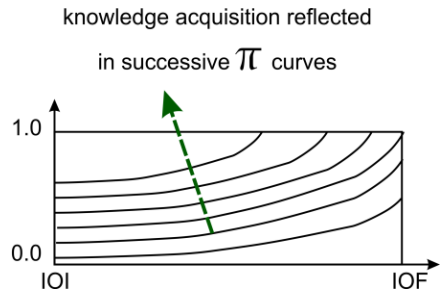

At any given state of knowledge the $\pi(I)$ curve will reflect its measure. There are various ways to reduce the $\pi(I)$ curve into a single numeric value to represent the state of knowledge. The simplest way is to track the area under it, since that area begins with zero (at total ignorance), and gradually increases with the process of knowledge acquisition until the state of total knowledge when its value is 1*(IOF-IOI). This reasoning leads to a definition of the level of knowledge associated with a given state:

$$ARK = \frac{\int_{I=IOI}^{I=IOF} \pi(I)dI}{IOF - IOI}$$

Where ARK stands for "*Available Relevant Knowledge.*" Clearly $0 \leq ARK \leq 1.0$

---

[1] If there are several values of $x$ corresponding to $\pi$ then the choice of $x^*$ will be made by some pre agreement.

As the knowledge-seeker acquires more relevant knowledge his or her ARK increases from total ignorance (ARK=0) to total knowledge (ARK=1), and the rate of progress depends strongly on the chosen values for IOI and IOF.

We can now define and quantify the knowledge that is still missing at any state of the knowledge acquisition process:

$$MARK = 1 - ARK$$

Where MARK stands for "*Missing Acquirable Relevant Knowledge*". Also $0 \leq MARK \leq 1.0$ only that MARK decreases from 1.0 to 0.0, opposite ARK.

### 3.2 Solution Signatures

The effort to resolve a given challenge is associated with a corresponding knowledge acquisition process, which can be measured and tracked using ARK or MARK. By plotting the decrease of MARK against the expenditure of cost or time (or any other appropriate resource), one creates a graphic signature for the challenge. It turns out that that signature is characteristic of the class of challenges to which the plotted challenge belongs, and that signature may be used for estimating the solution pathway of other challenges of the same class. [Samid-02, Samid-07b].

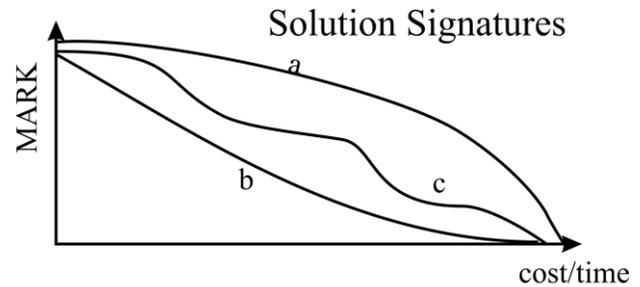

Challenges (a),(b),(c) represent three different classes of challenges, each represented by its distinct "Solution Finger Print" or "Solution Signature". The signature helps estimate the effort to solve other challenges of same class.

### 3.3 Knowledge Acquisition over a Discrete Set

A large class of challenges is represented as a selection dilemma among a discrete and final set of choices. We then define a reference challenge to pick the right choice among the competing candidates. A complete knowledge of relevance is the knowledge of the 'right' candidate. State of partial knowledge with respect to the relevant challenge is expressed by the finite probabilities associated with each candidate as to it being the right choice. The state of total knowledge is expressed by *p(r)=1, p(i ≠ r)=0* indicating that candidate *'r'* is the right one. The state of total ignorance is expressed as: *p(i)=1/n for i=1,2,3,....n* where n is the number of set choices.

We may build the expressions for this category of knowledge acquisition by deriving them from the continuous case described before.

- Transforming a Continuous Case Into a Finite Set.

- Ranked Choices
- Unranked Choices
- Tree Structures
- Composite List

**3.3.1 Transforming a Continuous Case Into a Finite Set**: The continuous case described in (3.1) may be smoothly transitioned to a discrete case by converting the probability function $p(x)$ into a finite histogram. The probability associated with each column (bar, rectangle) will be the value of $p(x)$ at the center point of that column. That probability will represent the chance for the particular column to be the 'right' choice. For an $n$-column histogram this process will convert the continuous $p(x)$ to a set of $n$ probability values, comprising $p_h(x)$. By making the histogram as fine as desired, one would achieve a 'smooth' transition as desired. While the continuous case envisioned an infinite range for the target variable $x$, the histogram version will work on the basis of a limited stretch for the $x$ variable from $L_x$=low boundary to $H_x$=high boundary. Since the gap $L_x$-$H_x$ can be made as large as desired, there is here too, a smooth transition, where the pay-off for the boundaries may be made as small as desired. We may set the size of the histogram column to be of width $h$, and thereby express the probability function via $n=(H_x-L_x)/h$ columns. Once the columns are so defined the case is fully discrete. One of the histogram columns represents the "right choice", and finding it becomes the reference challenge for the knowledge seeker.

One would naturally select the histogram column width, $h$, to correspond to the IOI (the smallest interval of interest, no need for a smaller interval). IOF will be equal to $f$ adjacent columns. Namely any statement that will place the 'right choice' column in a group of adjacent $f+1$ columns will be futile and useless. The $f$ adjacent histogram columns $i=1,2,3,....f$ will be columns of interest represented. One noteworthy distinction between the continuous case and the derived discrete case is that the search for $\pi$ will be done by skipping from one column to the next. For any given $p(x)$ that defines a histogram $p_h(x)$ the $\pi$ value per a given interval $i$ is given by:

$$\pi(i, p_h(x)) = [\sum_{j=k}^{j=k+i} p_h(j)]\, max_{for\ k=1\ to\ k=n-i}$$

And the value of knowledge at any given state of $p(x)$ is expressed as:

$$ARK = \frac{\sum_{I=IOI}^{I=IOF-1} \pi(I, p_h(x))}{IOF - IOI}$$

And corresponding to the continuous case we may write:

$$MARK = 1 - ARK$$

Both ARK and MARK range from "0" to "1" in opposite states. We shall define the nominal case as the one where the histogram as a whole represents the interval of interest, namely *IOF=f=n*, and *IOI=1*, shaping ARK into:

$$ARK = \frac{\sum_{I=1}^{I=n-1} \pi(I, p_h(x))}{n-1}$$

For the continuous case we consider an infinite *x* axis, which leads to the case of total ignorance defined as *p(x)→0* for all values of *x*, which leads to ARK=0. However, in the corresponding discrete case the probability of each histogram column amounts to *1/n ≥ 0* and the corresponding value of relevant knowledge for the nominal case becomes:

$$ARK(total\ ignorance) = \frac{\frac{1}{n} + \frac{2}{n} + \cdots \cdot \frac{n-1}{n}}{n-1} = 0.5$$

In other words, the level of total ignorance in the derived discrete formula is always 0.5 regardless of how *p(x)* is expressed through a histogram (how refined the column structure). The value of total knowledge is given by:

$$ARK(total\ knowledge) = \frac{\sum_{1}^{n-1}(1)}{n-1} = 1$$

**3.3.2. Ranked Choices:** In the above discrete case (derived from the continuous case) the *n* choices were placed in a fixed order (along the x axis). A related case is when the choices are ranked by some priority measure. If no choice has the same ranking as another then this case boils down to the one derived from the continuous situation, and is treated the same. In the case where some choices have not been ranked, or when two or more choices have equal ranking, we need to adjust our solution. The adjustment is based on handling unsorted items and tree-structures, and will be addressed in "composite sets".

**3.33. Unranked Choices:** This discrete case refers to choice candidates that are unsorted, unranked, and of equal footing as to their impact on the reference challenge. The question then is how to position the choices along an ordered line, to treat them as the fully ordered case. Now, since there is no order present, the probabilities themselves will serve for the purpose. The choices will be ordered so that for any *i < j (i,j=1,2,…n)* there exists *p(i) ≥ p(j)*.

**3.34 Tree Structures:** Knowledge cases if configured as discrete selections may be super-configured as several competing tree structures where some interesting attributes are present. Let a set of n choices contain a single 'right' choice, and let the reference challenge be the effort to find that right one. These *n* choices may be grouped into $n_1, n_2, ....n_k$ groups such that:

$$n_1 + n_2 + ...... n_k = n$$

The so defined groups may be further grouped $n'_1, n'_2.....n'_{k'}$ into super groups such that:

$$n'_1 + n'_2 + ...... n'_{k'} = k$$

and so on, until such step where the upper groups are combined into a 'root' and thereby the original n choices are strung into a tree structure. We designate the root as level 0 for the tree. The direct branches of the root are designated level 1, and the collection of *their* branches, level 2, etc. until level *d*, which is considered the depth of the tree.

One could define a reference challenge with respect to any level in any such tree. The challenge will be to identify the branch of the tree that contains the right choice. With respect to the root itself the knowledge is by definition complete: the right choice is contained in the root. We will also use the term "generation" to identify a level in the tree. A generation will comprise all the tree nodes that are of equal depth. We now define a generational challenge as the challenge to identify the generational node that contains the choice candidate in the original n choices. So that a tree defined as root branching into A, B, and C, and then A branching into A1 and A2, and B ranching into B1, B2, B3, and C branching into C1 and C2, the set of "leaves" on the tree is the set corresponding to the original 7 choices: A1, A2, B1, B2, B3, C1, C2. The 'orginal challenge' also called 'the bottom challenge' (or the second generational challenge) is to spot the right choice among these 7 candidates, and the challenge to find who among: A, B and C nodes contains the 'right' choice is the first generational challenge.

We now define a reference challenge for any given node in the tree. It would be to determine which of its branches contains the right choice, or if none does, to so determine.. The knowledge for that challenge will be captured by the likelihood of the branches plus the aggregate likelihood for the right original choice to be included outside the branches of that node. So if the k branches of a given node come with likelihoods of $p_1, p_2,...... p_k$, then the knowledge with respect to the reference challenge will be determined by: $p_1, p_2,...... p_k, (1-(p_1+ p_2+...... +p_k))$. This will insure that the knowledge level per each node will be determined over a probability distribution that sums up to 100%.

**Example:** a root is divided to A at 30% likelihood and B at 70% likelihood. The A node is divided to A1, A2 and A3 at rates: 10%, 15%, 5% respectively. The reference challenge is to determine where the right choice is: A1, A2, A3, or elsewhere. Hence ARK(10,15,5,70)=0.83.

**The case of total knowledge at the original challenge**: this situation will imply total knowledge for all the generational challenges. By contrast the case of total ignorance with respect to the original challenge may translate to some positive knowledge at the upper generational challenges. A total ignorance for the case of *n* candidates will imply *p(i)=1/n for i=1,2,....n*. If these n candidates are grouped such that *k < n* candidates form one group, and the other *(n-k)* candidates form another, then the generational knowledge will be computed on the basis of *(k/n and (n-k)/n)*. For all k values where $k \neq 0.5n$, there would appear some generational knowledge on top of zero knowledge at the original set. This attribute may govern the tree formation of the individual choices to improve the solution process.

**3.3.5 Composite List:** The list of candidates may be comprised of some which are well ordered, others that are of equal priority, and yet others for whom no order or priority attributes are available. Such a composite list will be processed as follows.

**Step 1:** place the well ordered candidates in the proper ordered list.
**Step 2:** order candidates of equal priority by placing them all in their priority ranking, and ordering them among themselves by descending probability. If two or more equal-rank order candidates have also the same probability then their order does not matter.
**Step 3:** the unordered candidates should be treated as in step 2, assuming they are last in the list.

**Example:** a candidate list comprised of 10 candidates with known probability ratings, is partially ordered, with some candidates sharing the same priority slot, as follows:

| CANDIDATE NAME | A1 | A2 | A3 | A4 | A5 | A6 | A7 | A8 | A9 | A10 |
|---|---|---|---|---|---|---|---|---|---|---|
| RANKING | 1 | 2 | 5 | 3 | 6 | 4 | 4 | ? | ? | 4 |
| PROBABILITY (%) | 26 | 12 | 8 | 15 | 5 | 2 | 3 | 20 | 6 | 3 |

We first execute step 1; ordering the well ordered elements: A1, A2, A4, A3, A5. Then we need to place all the items rank ordered at "4" in their proper location:
A1, A2, A4, [items ranked 4], A3, A5
These are items A6, A7 and A10. We rank these three according to their probability ratings: A7, A10, A6 or: A10, A7, A6, which is the same for our purpose. The ordered list now looks like:
A1, A2, A4, A7, A10, A6, A3, A5

Candidate A8, and A9 are unordered, so they are probability ranked: A8, A9 trailing the list. The final result is:

A1, A2, A4, A7, A10, A6, A3, A5, A8, A9

### 3.4 Higher Order Knowledge Equations:

The fundamental concept in the knowledge equations here is the full accounting of all the changes in the probabilities of the candidates but only to the extent that they impact the reference challenge. One defines a *'utility stretch'* and views the challenge in question as one which can be solved at varying degrees of utility. Any new knowledge is being judged as to its impact on the various utility levels for resolving that challenge. In the equation used so far the various impacts were strictly added, (then normalized). This method may be challenged by the logic that says: *the higher the utility, the more important the added knowledge*. So for interval $I < I'$, if in both cases the added knowledge is ΔK, then the added contribution to solving the challenge is greater for the I case. This logic may lead to defining a k-th order ARK formula, $ARK^K$:

$$ARK^k = \frac{\int_{I=IOI}^{I=IOF} \pi(I)(IOF - I)^k dI}{\int_{I=IOI}^{I=IOF} (IOF - I)^k dI}$$

And a corresponding formula for the discrete case.

### 3.5 Illustrations:

Let us consider a challenge to spot the right option among 10 candidates. As the knowledge seeker attacks this challenge the probabilities for each candidate change. The initial state of 'total ignorance' is characterized by p(i)=0.10 for i=1,2,..10. The probabilities distribution evolves, and the case terminates when p(4)=1.00 and p(i≠4)=0.0. We register 11 in-between states as follows:

| Candidate state | 1 | 2 | 3 | 4 | 5 | 6 | 7 | 8 | 9 | 10 | sum | **Shannon** | **MARK** |
|---|---|---|---|---|---|---|---|---|---|---|---|---|---|
| | | | | | 10 Candidates Selection Challenge | | | | | | | | |
| 1-start | 10.0 | 10.0 | 10.0 | 10.0 | 10.0 | 10.0 | 10.0 | 10.0 | 10.0 | 10.0 | 100 | **3.32** | **1.00** |
| 2 | 8.0 | 8.0 | 6.0 | 18.0 | 10.0 | 10.0 | 2.0 | 12.0 | 16.0 | 10.0 | 100 | **3.17** | **0.88** |
| 3 | 5.0 | 5.0 | 6.0 | 24.0 | 10.0 | 10.0 | 2.0 | 14.0 | 16.0 | 8.0 | 100 | **3.06** | **0.76** |
| 4 | 2.0 | 5.0 | 6.0 | 32.0 | 8.0 | 7.0 | 2.0 | 14.0 | 16.0 | 8.0 | 100 | **2.88** | **0.68** |
| 5 | 2.0 | 2.0 | 6.0 | 40.0 | 8.0 | 7.0 | 2.0 | 14.0 | 16.0 | 3.0 | 100 | **2.64** | **0.56** |
| 6 | 2.0 | 2.0 | 2.0 | 50.0 | 8.0 | 5.0 | 2.0 | 10.0 | 16.0 | 3.0 | 100 | **2.37** | **0.46** |
| 7 | 1.0 | 1.0 | 1.0 | 58.0 | 8.0 | 2.0 | 2.0 | 10.0 | 16.0 | 1.0 | 100 | **1.99** | **0.38** |
| 8 | 1.0 | 1.0 | 1.0 | 72.0 | 2.0 | 1.0 | 1.0 | 6.0 | 14.0 | 1.0 | 100 | **1.49** | **0.30** |

| 9 | 1.0 | 1.0 | 1.0 | 81.0 | 1.0 | 1.0 | 1.0 | 2.0 | 10.0 | 1.0 | 100 | **1.16** | **0.20** |
|---|---|---|---|---|---|---|---|---|---|---|---|---|---|
| 10 | 0.0 | 0.0 | 0.0 | 90.0 | 0.0 | 1.0 | 1.0 | 0.0 | 7.0 | 1.0 | 100 | **0.60** | **0.10** |
| 11 | 0.0 | 0.0 | 0.0 | 96.0 | 0.0 | 1.0 | 1.0 | 0.0 | 2.0 | 0.0 | 100 | **0.30** | **0.04** |
| 12 | 0.0 | 0.0 | 0.0 | 98.0 | 0.0 | 1.0 | 0.0 | 1.0 | 0.0 | 0.0 | 100 | **0.16** | **0.02** |
| 13-fini | 0.0 | 0.0 | 0.0 | 100.0 | 0.0 | 0.0 | 0.0 | 0.0 | 0.0 | 0.0 | 100 | **0.00** | **0.00** |

The knowledge acquisition progress was monitored by Shannon[2] and by MARK. The two metrics seem quite close:

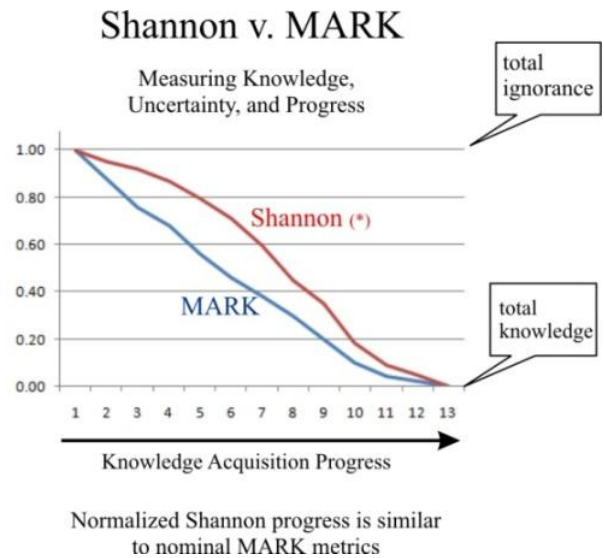

Normalized Shannon progress is similar to nominal MARK metrics

## 4.0 Comparative Analysis

The most common approach for estimating ignorance is analytic. It is based on an assumption that the data at hand follows a prescribed distribution. An analytical distribution has well established parameters to measure variability, which in turn is considered as the universal measure of uncertainty. The most common is the normal distribution, with its standard deviation serving as the highly popular metrics for variability. This approach is very well suited for handling variability due to errors in measurement. It's effective in answering the question whether two variables are statistically equal. The problem with this common approach is that it involves the arbitrary assumption of the data fitting into a given distribution, and also that standard deviation shrinks when the probability curve changes -- regardless of how "un-impactful" those changes are. This is the same weakness that is experienced by users of Shannon's entropy formula. Example: if a set of candidates comes with probabilities like: 40%, 33%, ....... 1%, 0.4%. then even a large percentage modification of the smallest value, say from 0.4% to 0.7%, will make no difference with respect to the difficulty of resolving the challenge of reference, finding the 'right' candidate. Yet, both the standard deviation, and Shannon's entropy will register a more advanced state for such a non consequential probability change.

---

[2] The graph depicted Shannon ratings were normalized to span 0-1, like MARK.

While standard deviation was migrated from error handling, Shannon's formula was brought about from communication theory. Its original use was to account for the regularities of the English language as compared to a random sequence of characters. Over the years its application range broadened -- to a large extent because of its very impressive name -- *entropy*, (reportedly based on a suggestion from Von Neumann explaining that the word entropy carries an air of mystery). It works well for the continuous case and for the discrete configuration. It is much less arbitrary than the analytic way, because it does not rely on assuming an analytical distribution. And it comes across as very convincing because as Shannon has shown [Shannon-49], it is the only analytic expression that would satisfy the somewhat arbitrary posted pre-requisites.

When one plots the ARK values over a normal (Gaussian) distribution, then these values increase when the IOF is increasing. In the graph below the ARK values per different spans (measured by standard deviations) are compared to the total area covered under the distribution. The ARK figures show a uniform increase under the area graph.

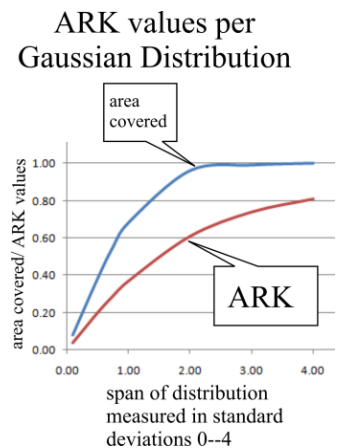

Both the standard deviation and Shannon's entropy may grow into infinity to express a growing amount of uncertainty, and in both cases there is no such a thing as 'total ignorance'. By contrast, the ARK and MARK formulas express 'total ignorance' as zero knowledge (per the reference challenge), which is defined as the state when the knowledge seeker has no ground to prefer one candidate over the other as to which one is the 'right' one.

The ARK and MARK expressions allow for setting the IOI and the IOF and thereby tailor the result to the case in point. If we consider *n* candidates, among which is the 'right' one then we can envision two extreme cases: (i) missing the 'right' one is a failure regardless which other candidate was selected, and (ii) the degree of failure depends on which alternative was selected. An example for the first case is: finding which piece of data was contaminated by a virus; an example for the second case is: which investment option will perform the best -- a second best selection will be better than the worst investment. For cases of the first type, one would set IOI=IOF, and for a case of the second type one would set IOF to a maximum stretch. In between cases will be marked with in-between IOF values. In other words, the MARK approach provides for tailoring the amount of relevant knowledge acquisition accounting. This advantage is not available for users of standard deviation, Shannon entropy, and its common replacements

## 5.0 Revisiting Risk, Opportunity, and Standard Deviation

The MARK formula is well positioned to serve as a basis for alternative expressions to capture risk, opportunity and variability. A brief discussion follows.

Risk is often associated with adversarial and harmful events, but in fact risk is a measure of the likelihood for harm and injury, not a measure of the harm itself. If we scheduled a root canal treatment then there is no risk involved only an inevitable bad experience. It follows then that if -- like in the case of the root canal -- we have complete knowledge about an impending hard, then we have eliminated the prospective risk. In other words, risk is a measure of our ignorance of the situation, and hence the *MARK* formula is applicable. Let $D_0$ be the measure of the damage from a given situation, should it be mishandled with total ignorance, and let $D_1$ be the measure of the damage if the situation would have been handled with total knowledge of the relevant factors. *D* may be measured in dollars or in any other metric. If we were to select the IOI and IOF in good measures then the computed expression $D_1+(D_0-D_1)*MARK$ will convey the expected damage under the situation of partial knowledge. The more we learn about the various prospects, the less the unpleasant surprise that we may sustain. It is similar and opposite with respect to opportunity. Let $O_0$ be the opportunity presented by a situation if it is handled with total ignorance, while $O_1$ will represent the opportunity if smartly exploited. *O* may be measured in dollars or otherwise. The expression $O_1-(O_1-O_0)*MARK$ will convey the de-facto opportunity based on how much we know about which moves are best. Lastly, ARK or MARK may serve as a generic means to measure variability of a parameter around its mean. The higher the MARK, the greater the dispersion. In that case one would set IOI to be very small and IOF to be very large.

## 6.0 Intractability Metrics

Prolonged and difficult projects and undertakings are often carried out for a long time without having any objective indication on whether there is any progress, or whether one is just spinning his wheels. In other words: *intractability erosion defies measurement*. Yet, progress depends directly on proper funding of those hard-to-measure endeavors. The significance and the nobility of the objective of the project is not a sufficient justification for committing resources to a particular way of going about it. If the difficult quest is not progressing towards a solution, then its investment is wasted, and the funds are being denied to a more productive alternative. Say then that prosperity and survival depend on our ability to measure progress in hard and difficult undertakings. A good metrics for missing relevant knowledge as it shrinks over time, is also a good basis to measure intractability and progress in eliminating it. The reason is that difficulty may always be associated with missing relevant knowledge. So if we can measure the missing relevant knowledge as it diminishes over time we can estimate how long and how costly it will be to solve the challenge as a whole. Samid has used this approach to appraise the effort required to achieve various research and development objectives (Samid-02).

If at time point 1 the residual missing acquirable relevant knowledge (residual-MARK), or MARK) is $M_1$, and at point 2 the MARK value is $M_2$, and if to proceed from state 1 to state 2 one expended $T$ resources (time, money or otherwise), then the expression:

$$\frac{T}{M_1 - M_2}$$

reflects the apparent intractability of the project between these states (measured by the units of resource T). The expression $\frac{dT}{dM}$ will reflect the local intractability, and:

$$\int_{M=1}^{M=0} (\frac{dT}{dM}) dM$$

will reflect the overall intractability of the project as a whole. Difficult undertaking can then be categorized by the MARK over time or MARK-over-cost curves to help sort out any pending intractability, and provide useful guidance for optimal allocation of our scarce resources as we quest to apply them towards our most intractable challenges.

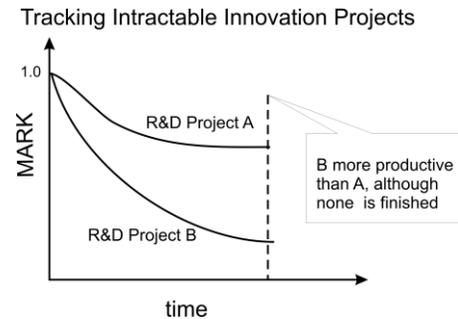

## 7.0 Perspective

We are in the *Age of Knowledge*, our survival and prosperity depends on our ability to direct our resources to acquiring relevant knowledge. And thus, it is inherently important for us to be able to measure the knowledge acquisition process, and the amount of relevant knowledge still missing. So important is it that however daunting the task, however preliminary this effort, it may be a worthy one. We cannot offer a rational strategy for solving our mounting problems without being able to credibly appraise the required knowledge acquisition effort . And that appraisal cannot take place without a tool and a methodology to deduce its conclusions with science and objectivity. Mindful that hitherto this appraisal was the domain of art and intuition one should read and review this article in the light of being but an embryonic step.